\documentclass [preprint,aps] {revtex4}
\usepackage{graphicx}
\begin{document}
\draft
\title{Scaling Properties of Long-Range Correlated Noisy Signals}
\author{Anna Carbone and Giuliano Castelli}
\address{Dipartimento di Fisica, Politecnico di Torino,
Corso Duca degli Abruzzi 24, I-10129 Torino, Italy\\
and Istituto Nazionale di Fisica
della Materia, Unit\`a del Politecnico di Torino}
\date{\today}

\begin{abstract}

The Hurst coefficient $H$ of a  stochastic fractal signal is
estimated using the function
$\sigma_{MA}^2=\frac{1}{N_{max}-n}\sum_{i=n}^{N_{max}}
[y(i)-\widetilde{y}_n(i)]^2$, where $\widetilde{y}_n(i)$ is
defined as $1/n \sum_{k=0}^{n-1} y(i-k)$,  $n$ is the dimension of
moving average box and $N_{max}$ is the dimension of the
stochastic series. The ability to capture scaling properties by
$\sigma_{MA}^2$  can be understood by observing that the function
$C_n(i)= y(i)-\widetilde{y}_n(i)$ generates a sequence of random
clusters having power-law probability distribution of the
amplitude and of the lifetime, with exponents equal to the fractal
dimension $D$ of the stochastic series.

\end{abstract}
\maketitle

Power-law distributions are important manifestations of scale
invariance as observed in fractal, percolating structures and
self-organized dynamical systems. The scaling exponents of power
law statistics are indeed related to the universality class of the
system and thus are helpful to understand the fundamental
processes ruling the dynamics of complex systems. The scaling
parameters have been practically deployed in fields as different
as biology, geophysics, solar physics, social science. When random
time records should be analyzed the Hurst exponent $H$, which is
related to the fractal dimension of the time series by $D=2-H$, is
usually evaluated. For example, healthy and sick heart beat rate
can be distinguished on the basis of the value of $H$. Financial
series with degree of persistence higher than that of the price
series have been encountered.
  A
number of frequency, time and, even, {\em integrated} domain
approaches have been thus developed to gain as accurate as
possible estimate of these exponents. Such procedures generally
consist in calculating appropriate statistical functions from the
whole signal: in the time domain,  Detrended Fluctuation Analysis
(DFA) and Rescaled Range Analysis (R/S) are the most popular
scaling methods to extract power-law correlation exponents from
the random signals $y(i)$
~\cite{Feder,Mantegna,Bak,Sethna,Vandewalle,KunHu,Rangarajan,Heneghan,Alessio,Lux,Ivanov}.
\par
   A
generalized variance $\sigma_{MA}^2$ of the noisy signal $y(i)$
with respect to $\widetilde{y}_n(i)$, defined by:
\begin{equation}
\sigma_{MA}^2 = {{\frac{1}{N_{max}-n}
\sum_{i=n}^{N_{max}}{C_n(i)^2} }} , \label{sigmama}
\end{equation}

with:

\begin{equation}
 C_n(i)= y(i)-\widetilde{y}_n(i)   ,
 \label{Ci}
\end{equation}

 has been introduced in \cite{Alessio}.
In Eq.(\ref{sigmama}), $N_{max}$ is the size of the series,
$\widetilde{y}_n(i)$ is the moving average  defined by
$\widetilde{y}_n(i)=\frac{1}{n}\sum_{k=0}^{n-1} y(i-k)$ and $n$ is
the moving average window.
 \par The function ${\sigma_{MA}}$ corresponding to
each $\widetilde{y}_n(i)$ has been calculated for random series
$y(i)$ with different Hurst exponent $H$. The values of
${\sigma_{MA}}$ are plotted as a function of $n$ on log-log axes
shown in Fig.[\ref{sigma}], for a series with $N_{max}=2^{19}$ and
$10\leq n\leq10^4$. The most remarkable property of the curves
plotted in Fig.[\ref{sigma}] is their power-law dependence on $n$,
i.e.:

\begin{equation}
\sigma_{MA}  \approx n^{H} .
 \label{sigma}
\end{equation}

The structure of the algorithm based on the $\sigma_{MA}$ function
is analogous to the DFA technique, but a higher accuracy has been
observed. This may be due to the better smogthing and detrending
action of the moving average with respect to the linear or to the
polynomial fit. Furthermore, the moving average filter dynamically
detrends the series. Every time the discrete index $i$ increases
by a unity, the box window $n$ switches its position accordingly,
allowing also for online applications of the present technique.
 In order to gain  a deeper
 insight of the function ${\sigma_{MA}}$, we will show that $C_n(i)$ generates, for each $\widetilde{y}_n(i)$, a sequence of
clusters with amplitude and lifetime distributed as power-laws.
Furthermore the exponent of these power-laws is equal to tte
fractal dimension $D=2-H$ of the time series. These properties
allow to understand the ability of ${\sigma_{MA}}$ to estimate the
Hurst exponent $H$ of the stochastic series.
\par
\begin{figure}
\leftline{\includegraphics[width=1\columnwidth,angle=-90]{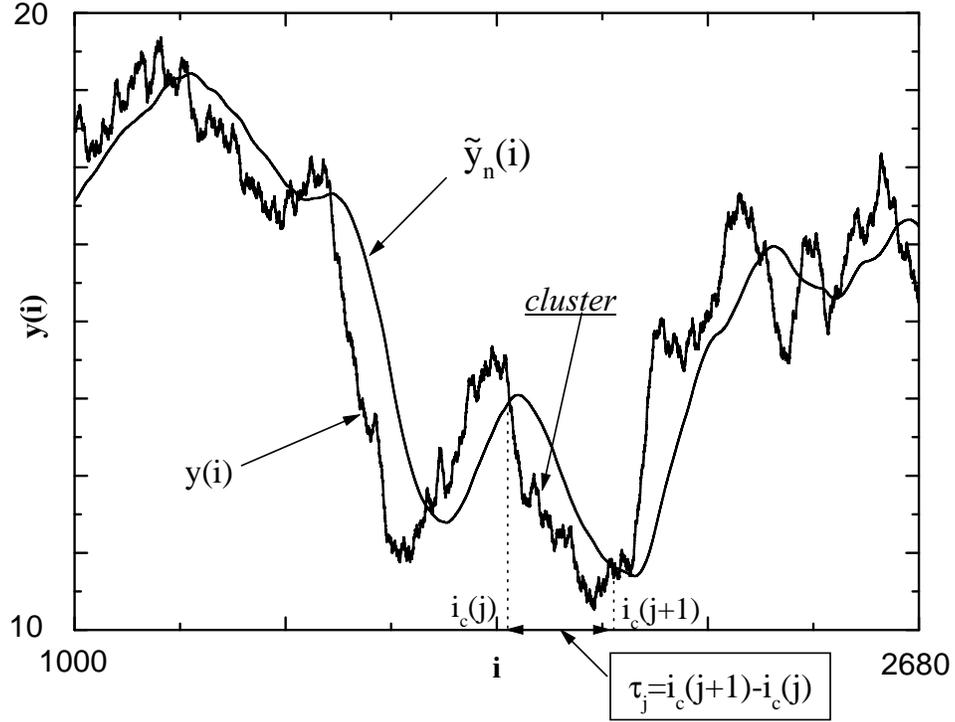}}
\vspace{-3cm} \caption{\label{cluster} Stochastic series $y_n(i)$
obtained by the Random Midpoint Displacement algorithm with
$H=0.3$. The size of the series is $N_{max}=2^{19}$. The moving
average $\widetilde{y}_n(i)$, with box dimension  $n=30$, is also
shown. The total length of the segment and the time interval
between two subsequent crossing points represent respectively the
size and the duration (lifetime) of the cluster according to the
definition given in the text (Eqs.(\ref{l},\ref{tau})).}

\end{figure}

Consider thus the sub-set of $y(i)$ corresponding to the region
delimited by two consecutive intersections between $y(i)$ and
$\widetilde{y}_n(i)$. Let us refer to such regions as {\em
cluster}. In Fig.[\ref{cluster}], one of these clusters is shown.
For each cluster $j$, the {\em size} $l_j$ :

\begin{equation}
\label{l} l_j={{\sum_{i=i_c(j)}^{i_c(j+1)}{y(i)} }}
\end{equation}

and the duration $ \tau_j $:

\begin{equation}
\label{tau}
 \tau_j = i_c(j+1)- i_c(j) .
\end{equation}

can be defined. In the previous relationships, $i_c(j)$ and
$i_c(j+1)$ are the values taken by the index $i$ in correspondence
of two subsequent intersections between $\widetilde{y}_n(i)$ and
$y(i)$. The size $l_j$ and the duration $\tau_j$ of the clusters
are shown in Fig.(\ref{cluster}) for a series with
$N_{max}=2^{19}$, $H=5.3$ and $n=30$.

\par
\begin{figure}
\leftline{\includegraphics[width=1\columnwidth,angle=-90]{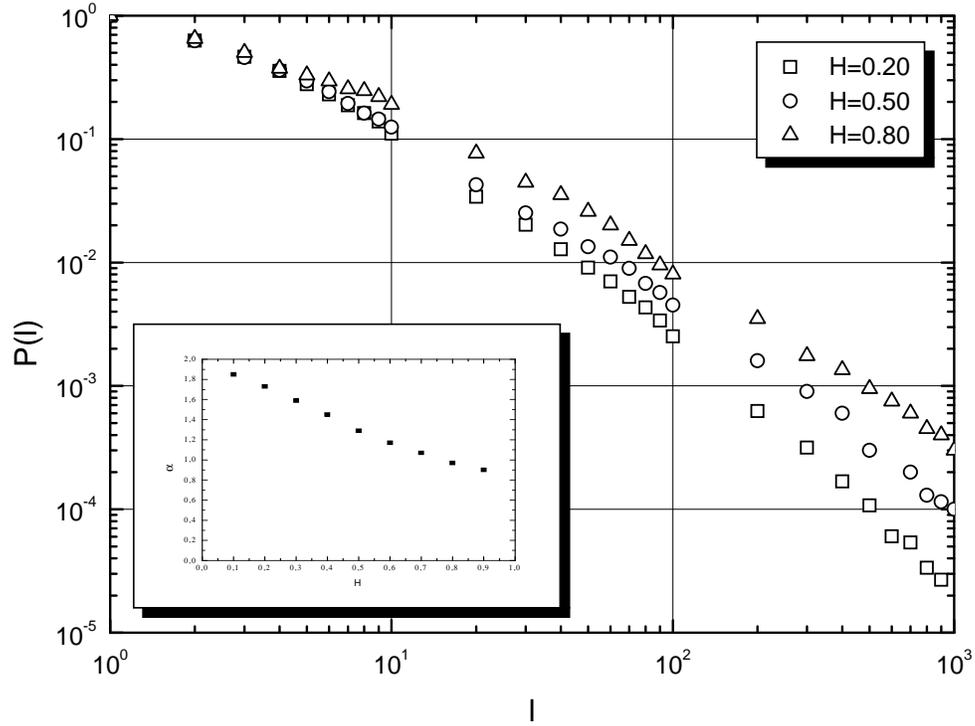}}
\vspace{-3cm} \caption {\label{Dlf} Log-log plot of  distribution
$P(l)$ of the cluster size $l$ for series having different $H$
(respectively $H=0.2$, $H=0.5$ and $H=0.8$). The clusters are
obtained by the intersection of the series with a moving average
with a  window amplitude $n=60$. A power-law dependence
$P(l)\approx l^{-\alpha}$ has been observed. It is
worth-mentioning that the exponent $\alpha$ depends on the $Hurst$
exponent of the series, but is independent of the series dimension
$N_{max}$ and of the moving average box $n$.}
\end{figure}

The probability distribution $P(z)$ of the cluster sizes has been
determined by counting the number of clusters with length $l$.
Fig.[\ref{Dlf}] shows the log-log plot of the distribution $P(l)$
for series having different $H$ (respectively $H=0.2$, $H=0.5$ and
$H=0.8$). The clusters are obtained by the intersection of the
series with a moving average with a  window amplitude $n=60$. The
curves are consistent with a straight line, indicating a power-law
behavior:

\begin{equation}\label{Ps}
  P(l) \approx l^{-\alpha}
\end{equation}

The linear behavior over two decades indicates a scaling
distribution of clusters. The exponent $\alpha$ has been plotted
against $H$ can the inset of Fig.[\ref{Dlf}].

\par
\begin{figure}
\leftline{\includegraphics[width=1\columnwidth,angle=-90]{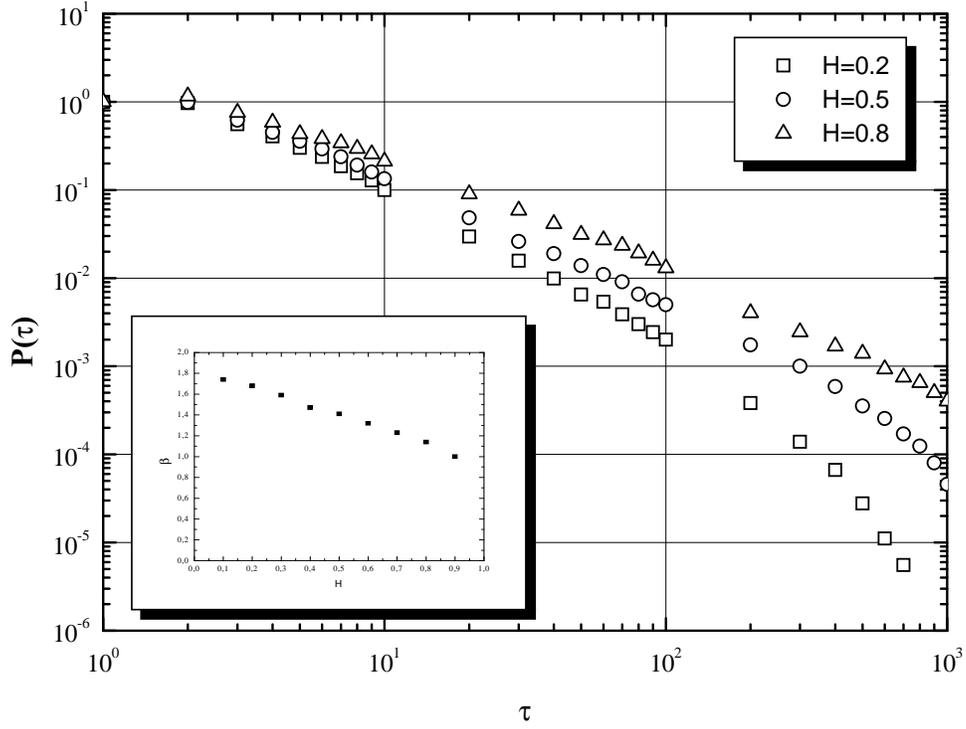}}
\vspace{-3cm} \caption {\label{Dtauf} Log-log plot of the
distribution $P(\tau)$ of the cluster lifetime $\tau$ for series
having different $H$ (respectively $H=0.2$, $H=0.5$ and
$H=0.8$).The clusters are obtained by the intersection of the
series with a moving average with a  window amplitude $n=60$. The
lifetime of the cluster corresponds to the time interval between
two consecutive intersection points, as defined by the
Eq.(\ref{tau}). A power-law dependence $P(\tau)=\tau^{-\beta}$ has
been observed. It is worth-mentioning that the exponent $\beta$
depends on the $Hurst$ exponent of the series,  but is independent
of the dimension $N_{max}$ and of the moving average box $n$. }

\end{figure}

The probability distribution $P(\tau)$ of the cluster lifetime has
been determined by counting the number of clusters with lifetime
$\tau$. Fig.[\ref{Dtauf}] shows the log-log plot of the
distribution $P(\tau)$  for series having different $H$
(respectively $H=0.2$, $H=0.4$ and $H=0.8$). The clusters are
obtained by the intersection of the series with a moving average
with a  window amplitude $n=60$. This leads to another line
indicating a distribution of lifetimes of the form:

\begin{equation}\label{Ptau}
P(\tau) \approx \tau^{-\beta}   .
\end{equation}

The values of  $\alpha$ and $\beta$,  for $H$ varying from $0.10$
to $0.90$, are reported in Table (1).
 The distributions  $P(l)$ and $P(\tau)$ have
been calculated for a wide range of values of  $N_{max}$ and $n$ (
$2^{18}< N_{max}<2^{21} $ and $60<n<1000$), the exponents $\alpha$
and $\beta$ have been found to be independent of $N_{max}$ and
$n$. It can be concluded that $\alpha$ and $\beta$ vary as $2-H$,
i.e. as the fractal dimension $D$ of the noisy signal.
\par These results can be
understood keeping in mind the box-counting method to estimate the
fractal dimension of a random signal is recalled (Ch.10 of
\cite{Feder}). The stochastic signal is covered with boxes of
width $b \tau$ (in time) and of length $b a$ (in amplitude), with
$\tau \cdot a$ the minimum box. The box dimension $D_B$ is related
to the number $N(b;a,\tau)$ of the boxes needed to cover the
record by the relationship: $N(b;a,\tau)\approx b^{- D_B}$, where
$D_B=2-H$ for self-affine records.
\par
 We have demonstrated that the Hurst exponent $H$ of
 long-range correlated stochastic series
$y(i)$ can be
 calculated
 using the function $\sigma_{MA}$ defined by the
Eq.(\ref{sigmama}). We have found indeed the remarkable result
that the function $\sigma_{MA}$ varies as a power-law of the
amplitude $n$ of the moving average box. This work shed more light
on the results of the papers \cite{Vandewalle,Alessio} and
reinforce the general idea that a deeper physical insight may add
new perspectives to the practical methods of financial analysis.
\par
These results, appeared in very preliminary form in
\cite{Alessio}, have been here validated in terms of the
statistical properties of the random clusters tracked by
$\widetilde{y}_n(i)$ over the random time series. Furthermore, the
power-law behavior of the distributions of cluster size and
lifetime, just as expected from general arguments about dynamical
system, clarifies the intrinsic capability of the function
$\sigma_{MA}$ to capture scaling exponents of the series.
Furthermore, the exponents $\alpha$ and $\beta$ of
Eqs.(\ref{Ps},\ref{Ptau}) are insensitive to the parameters
characterizing $\widetilde{y}_n(i)$ and  $y(i)$. This robustness
is essential to the assessment of the proposed scaling technique.
Last but not least, the $\sigma_{MA}$ algorithm yields in higher
accuracy, speed of execution  and  possibility of online
application.

\begin{figure}[c]
\leftline{\includegraphics[width=1\columnwidth,angle=-90]{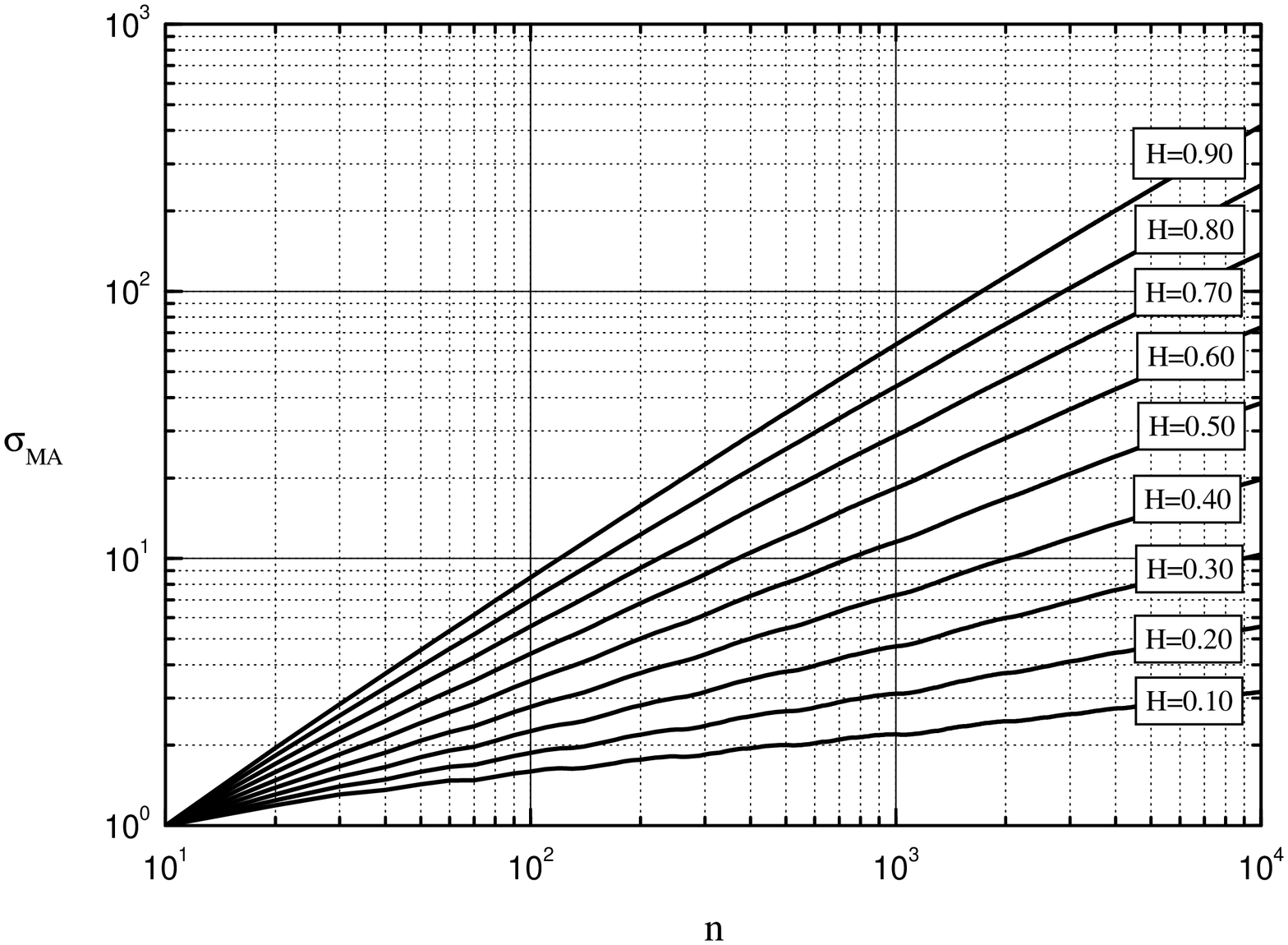}}
\vspace{-3cm} \caption{\label{sigma} Log-log plot of the function
$\sigma_{MA}$ vs. the moving average box $n$. The curves have been
obtained using the computational algorithm described in the text.
The curves refers to artificially generated series having
$N_{max}=2^{19}$ and Hurst coefficient $H$ varying between 0.10
and 0.90. A power-law dependence ($\sigma_{MA} \propto n^{H}$) is
found. The results are independent of the series dimension
$N_{max}$.}
\end{figure}

\end{document}